\documentclass[twocolumn]{aastex7}
\usepackage{float}
\usepackage{hyperref}
\usepackage{amsmath}
\usepackage{bm}

\begin{document}

\title{Ionization Rate of Interstellar Neutral Helium from New Horizons/SWAP Observations}

\author[orcid=0000-0002-8581-9386]{Małgorzata Antonik}
\affiliation{Space Research Centre PAS (CBK PAN), Bartycka 18a, 00-716 Warsaw, Poland}
\email[show]{mantonik@cbk.waw.pl}  

\author[orcid=0000-0002-9033-0809]{Paweł Swaczyna} 
\affiliation{Space Research Centre PAS (CBK PAN), Bartycka 18a, 00-716 Warsaw, Poland}
\email{pswaczyna@cbk.waw.pl}  

\author[orcid=0000-0001-6160-1158]{David J. McComas}
\affiliation{Department of Astrophysical Sciences, Princeton University, Princeton, NJ 08544, USA}
\email{dmccomas@princeton.edu}  

\author[orcid=0000-0003-2297-3922]{Heather A. Elliott}
\affiliation{Southwest Research Institute, San Antonio, TX 78228, USA}
\email{heather.elliott@swri.org}  

\author[orcid=0000-0003-3957-2359]{Maciej Bzowski} 
\affiliation{Space Research Centre PAS (CBK PAN), Bartycka 18a, 00-716 Warsaw, Poland}
\email{bzowskicbk.waw.pl}

\newpage
\begin{abstract}

Interstellar neutral (ISN) atoms enable studies of the physical conditions in the local interstellar medium surrounding the heliosphere. ISN helium, which is the most abundant species at 1 au, is directly observed by space missions, such as Interstellar Boundary Explorer (IBEX). However, some of these atoms are ionized by solar ultraviolet radiation before reaching 1 au, producing pickup ions (PUIs). A recent analysis of IBEX  data suggests that the helium photoionization rates predicted by models are underestimated by up to 40\%. The Solar Wind Around Pluto (SWAP) instrument on board New Horizons enables the study of PUIs giving complementary insight into the other side of the ionization process. Our goal is to verify this increased helium ionization by determining the ionization rate of ISN helium in the heliosphere based on the SWAP observations of helium PUIs. For this purpose, we analyze SWAP data collected between 2012 and 2022, at distances 22 to 54 au from the Sun. We develop a new method for fitting model distribution functions to the observational data using the maximum likelihood method. Our approach accounts for the spacecraft’s rotation and the SWAP response function, which depends on both energy and inflow direction. We estimate SWAP’s efficiency for helium relative to that for hydrogen and determine the ISN helium ionization rate. We find that the photoionization rate obtained from the SWAP observations is $\sim$43\% larger than the rates predicted by models, confirming the IBEX results. 

\end{abstract}


\section{Introduction} \label{sec:introduction}

The solar wind is a continuous stream of magnetized plasma emitted from the solar corona, propagating outward through the Solar System \citep{1958ApJ...128..664P}. It is mainly composed of electrons, protons ($\mathrm{H^+}$), and alpha particles ($\mathrm{He^{2+}}$),  accelerated to supersonic speed. The typical alpha to proton density ratio is typically a few percent \citep{2007ApJ...660..901K, elliott_determining_2018}. The abundance of singly ionized helium ($\mathrm{He^+}$) is much lower and is strongly influenced by solar activity. Under quiet solar conditions, its density is typically about $10^{-6}$ times that of alpha particles \citep{1966ApJ...144..244T, 1968SoPh....5..410K}, but it can increase by several orders of magnitude when the cool prominence material is embedded within the coronal mass ejection \citep{1999GeoRL..26..157G, 1999GeoRL..26..161S}. In addition, trace amounts of heavier ions, such as highly ionized oxygen, carbon, and iron, are present in the solar wind \citep{2000RvGeo..38..247B, 2007A&A...471..315B, 2000JGR...10527217V, 2007SSRv..130..139G, gilbert_first_2012}.

The solar wind and interplanetary magnetic field are responsible for the existence of the heliosphere, a bubble surrounding the entire Solar System moving through interstellar matter and extending to a hundred astronomical units \citep{1961ApJ...134...20P, 2019NatAs...3.1007B}. At its boundary, called the heliopause, the solar wind collides with partially ionized plasma of the very local interstellar medium (VLISM) \citep{1986AdSpR...6b...5B}. The VLISM plasma is deflected and flows around the heliopause. However, interstellar neutral (ISN) atoms cross the heliopause and continue their journey towards the Sun \citep{1971A&A....14..263F, 1972NASSP.308..609A, 1975Natur.254..202W}. Along the way, their trajectories are deflected by the solar gravity and radiation pressure. Additionally, as a result of photoionization, electron impact ionization, and charge exchange, some of them turn into pickup ions (PUIs) \citep{1972NASSP.308..609A, 1985Natur.318..426M}. Ions gyrate around magnetic field lines frozen into the solar wind plasma, and their pitch angle quickly becomes isotropic. PUIs are not thermalized with the core solar wind and can therefore be studied as a separate population. PUIs co-travel outward through the Solar System as a part of the solar wind plasma \citep{1970A&A.....4..280B, 1976JGR....81.1247V}. 

We obtain information about the core solar wind, PUIs, and ISN atoms thanks to numerous space missions. Based on these observations, it is possible to determine the physical conditions in the VLISM \citep{2004A&A...426..897M, 2019ApJ...882...60B, swaczyna_interstellar_2023}. ISN helium atoms are less affected in the heliosphere than ISN hydrogen, which is why they are mainly used to derive the VLISM flow parameters. Examples of missions enabling direct observations of the ISN helium atoms are Ulysses/GAS \citep{1992A&AS...92..333W, 1993AdSpR..13f.121W} and IBEX \citep{2009SSRv..146...11M}. Based on these observations, parameters such as ISN helium inflow velocity vector, Mach number, and direction were determined \citep{2004A&A...426..835W, bzowski_neutral_2014, 2015ApJS..220...28B,wood_revisiting_2015, 2015ApJS..220...25S, swaczyna_very_2022, swaczyna_interstellar_2023}.

The ISN helium density is usually assumed to be homogeneous and invariable at the boundaries of the heliosphere \citep{2004A&A...426..845G}. However, the flux of ISN atoms is reduced by ionization processes. The main ISN helium ionization process is photoionization by ultraviolet radiation from the Sun \citep{1995A&A...296..248R, 1996SSRv...78...73R, 2013ccfu.book...67B, 2016MNRAS.458.3691S, 2019ApJ...872...57S, 2020ApJ...897..179S}. At a distance of 1 au, electron impact ionization is also significant, contributing about 15\% of the total helium ionization, while charge exchange accounts for 3\% \citep{2019ApJ...872...57S}. Both photoionization and charge exchange ionization rates decrease with the square of the distance from the Sun, whereas the electron impact ionization rate decreases more rapidly, making this process negligible in the outer heliosphere \citep{1989A&A...224..290R, 2013A&A...557A..50B}. As a result of ionization, a singly-ionized $\mathrm{He^{+}}$ PUI is formed. Doubly ionized $\mathrm{He^{2+}}$ PUIs can be created by double charge exchange with the solar wind $\mathrm{He^{2+}}$ ions, but their population is significantly smaller \citep{1998A&A...334..337R, 2017ApJ...840...75S}. Nevertheless, the observations of $\mathrm{He^{2+}}$ and $\mathrm{He^{+}}$ PUI from Ulysses were used to determine the ISN helium density \citep{2004A&A...426..845G, 2004AdSpR..34...53G}. The local fluxes of ISN helium atoms and $\mathrm{He^{+}}$ PUIs depend on both the ISN helium density at the entrance to the heliosphere and their ionization rate. While both populations are proportional to the ISN helium density in the entrance, the photoionization decreases the ISN atoms flux and increases the $\mathrm{He^{+}}$ PUIs flux.

Recently, \cite{swaczyna_very_2022} analyzed IBEX-Lo data of ISN helium atoms over a full solar cycle and observed modulations of the ISN helium fluxes at 1 au. The results were compared with simulations from the Warsaw Test Particle Model (WTPM) \citep{2015ApJS..220...27S}.  They found that ionization rates calculated based on a series of observations of the solar EUV spectrum by TIMED (Thermosphere Ionosphere Mesosphere Energetics and Dynamics) \citep{2005JGRA..110.1312W, 2018SoPh..293...76W}, correlated with the solar 10.7 flux \citep{2014arXiv1411.4826S,2020ApJ...897..179S} may be underestimated by up to $\sim$40\%. The objective of our study is to verify this result by determining the helium ionization rate based on observations of helium PUIs. Alternatively, the observed long-scale modulation of ISN helium flux might also be attributed to the variation of ISN helium density over time.

In this study, we utilize data from the New Horizons spacecraft \citep{stern_new_2008}, a space probe launched in 2006. Its primary mission, completed successfully in 2015, was to perform a flyby study of Pluto and its moons. However, the spacecraft continues to provide valuable scientific data as it follows its trajectory, escaping the Solar System near the ecliptic plane. The onboard Solar Wind Around Pluto (SWAP) electrostatic instrument \citep{2008SSRv..140..261M} detects hydrogen and helium ions in the solar wind. SWAP for the first time provides the capability to investigate the PUIs in the outer heliosphere \citep{2010JGRA..115.3102M, randol_observations_2012, randol_interstellar_2013, 2017ApJS..233....8M, 2021ApJS..254...19M, 2022ApJ...934..147M, 2025ApJ...980..154M}. In particular, it is the only instrument measuring helium PUIs far beyond the ionization cavity \citep{2017ApJS..233....8M}, enabling study of the helium ionization rate based on the detection of the product of this ionization process. Furthermore, distant vantage point limits importance of the highly uncertain electron impact ionization, which is important at 1 au \citep{1989A&A...224..290R}. The core solar wind protons and alpha particles observed by SWAP are analyzed in \cite{elliott_new_2016, elliott_determining_2018, elliott_slowing_2019}. Moreover, SWAP makes it possible to study PUI acceleration at interplanetary shocks \citep{2018PhRvL.121g5102Z, 2024ApJ...960...35S, 2025ApJ...984...11S}, and identify the core solar wind $\mathrm{He^{+}}$ ions \citep{swaczyna_he_2019}. 

The study of helium PUIs observed by New Horizons/SWAP provides the basis for determining the ionization rate of ISN helium in the heliosphere, which is the central objective of this work. Section~\ref{sec:observations} describes the data selection of the SWAP observations used in the analysis. Section~\ref{sec:methods} introduces the methods applied in our study, including derivation of the SWAP energy–angle response function and determination of the solar wind and PUI parameters. Section~\ref{sec:results} presents the obtained results, while Section~\ref{sec:discussion} discusses possible uncertainties and an alternative hypothesis. Finally, Section~\ref{sec:summary} summarizes the main conclusions of this study.

\section{SWAP Observations} \label{sec:observations}

SWAP is designed to simultaneously measure solar wind ions and PUIs. It has a very large field of view (FOV) $\mathrm{276^\circ \times 10^\circ}$, aligned with the spacecraft’s antenna. SWAP collects the data in interlaced coarse and fine scanning modes, which are collected within 64 seconds and represent a 64-second snapshot \citep{elliott_new_2016}. Coarse scanning observations provide numbers of counts in 64 logarithmically spaced energy bins from approximately 0.024 to 7.6 $\mathrm{keV \, q^{-1}}$ until 2021 August 9, and from 0.022 to 5.0 $\mathrm{keV \, q^{-1}}$ after that date \citep{2022ApJ...934..147M}. The energy bin width is $\Delta E/E \approx 8.5\%$ FWHM. Unfortunately, the cadence at which these snapshots are recorded varies based on the available data downlink, but often ranges between 1 per hour and 1 per 10 minutes. Therefore, we use histogram-type spectra, combining multiple coarse scans. SWAP measures the combined energy-per-charge spectrum of solar wind ions and PUIs, but does not identify ion species. However, due to differences in masses and charges, the maxima of solar wind species are observed in different parts of the energy-per-charge spectrum, and therefore, can be distinguished from one another and analyzed individually.

We analyze data from almost ten years of coarse scanning observations (January 2012 – July 2022) with a time resolution of one day. Since February 19, 2021, spectra with a higher time resolution of approximately 30 minutes are available \citep{2022ApJ...934..147M, 2025ApJ...980..154M}. To maintain data consistency in our study, we aggregate the high resolution histograms to obtain a daily average spectrum. We select the analyzed time period based on the availability of mostly continuous data from SWAP \citep{2017ApJS..233....8M}.New Horizons is moving away from the Sun while remaining close to the ecliptic plane. We only use observations collected in the spinning mode (i.e. excluding 3-axis periods) during which the spacecraft rotates about an axis aligned with the antenna and the center of the SWAP FOV. The antenna points toward Earth for communication. Even during hibernation mode, the angular distance between the antenna’s actual orientation and the Earth’s position is negligible at large heliocentric distances. The spacecraft rotates about its axis at roughly 5 RPM (revolutions per minute), a configuration that allows the SWAP instrument to detect PUIs incoming from various angles. During the analyzed time period, the spacecraft was at distances ranging from approximately 22 to 54 au from the Sun, heading within the ecliptic plane roughly 30$^\circ$ away from the ISN inflow direction.

We further cull the data based on several criteria similar to those described in \cite{swaczyna_density_2020} and \cite{2021ApJS..254...19M}:
\begin{itemize}
    \item the number of coarse sweeps in the averaged spectrum has to exceed 1230 (the mode was 1284 \citep{randol_interstellar_2013}),
    \item there must be more than 12 measurements of solar wind properties from the fine sweeps \citep{elliott_new_2016, elliott_determining_2018, elliott_slowing_2019},
    \item the daily variation of the solar wind in the analyzed day from these measurements must be less than 1\% to ensure that the solar wind conditions did not vary too much during the day long histogram collection interval,
    \item we exclude spectra with significant PUI tails which indicate shock processing \citep{2018PhRvL.121g5102Z}, requiring that the count rate in the energy bins above five times the proton peak energy remain below one count per second (as previously used by \citet{2021ApJS..254...19M}),
    \item the background from penetrating particles \citep{randol_observations_2012} calculated based on the average of the count rate in the first four energy bins cannot exceed 0.15 s$^{-1}$.
\end{itemize}
Based on these criteria, we selected a total of 2384 observation days for our analysis.

\section{Methods} \label{sec:methods}

To achieve the primary goal of this study, which is to determine the helium ionization rate from SWAP observations of helium PUIs, we need to correctly interpret observations of helium ions. In this section, we describe the methodology used to process and analyze the SWAP data. We first focus on deriving the SWAP energy–angle response function for rotation-average FOV (Section \ref{subsec:response}). Next, we present the distribution functions adopted for both the core solar wind and the PUIs (Section \ref{subsec:distributions}), followed by a description of the fitting procedure used to determine the physical parameters of these populations (Section \ref{subsec:models}). Finally, we outline the criteria applied to select energy channels suitable for reliable model fitting (Section \ref{subsec:selection}).

\subsection{SWAP energy-angle response function} \label{subsec:response}

The response function of the SWAP instrument depends on the energy per charge and inflow direction. Due to the top-hat design of the instrument, there is a strong correlation of the energy response with the elevation angle ($\theta$) between the incoming particle direction and the FOV plane perpendicular to the instrument's approximate symmetry axis. The complementing azimuthal angle ($\phi$) describes the angle within this plane. The determination of the relationship was possible thanks to laboratory calibration and numerical simulations \citep{nicolaou_properties_2014, elliott_new_2016}. We use a table of simulated and scaled energy-angle response arrays ($E_{\mathrm{beam}}/E_{\mathrm{step}}$ vs. $\theta$) for each energy step, as provided by \cite{elliott_new_2016}. Since these arrays did not vary systematically across different energy steps, we construct a single, averaged energy-angle response function $R_{\mathrm{E,\theta}}(E/E_{\mathrm{step}}, \theta)$ using the same energy and angular resolution. 

The SWAP response function is essential to estimate the model count numbers for a given energy step $C(E_{\mathrm{step}})$ measured by the instrument:
\begin{equation}\label{eq:int4}
\begin{split}
   & C(E_{\mathrm{step}}) = \Delta t \, v_{\mathrm{c}} G^v(E_{\mathrm{step}}) g(t) \frac{1}{2\pi} \int_{0}^{2\pi}d\psi \int_{-\pi}^{\pi}d\phi  \\ & \int_{-\frac{\pi}{2}}^{\frac{\pi}{2}} \mathrm{cos} \, \theta \, d\theta \int_{v_{min}}^{v_{max}}dv \, f(\vec{v}) R_{\mathrm{E,\theta}}(E/E_{\mathrm{step}}, \theta) R_{\phi}(\phi) v^3,
\end{split}
\end{equation}
where $\Delta t$ is the observation time per energy bin, $v_{\mathrm{c}}$ is the center speed for given energy step, $G^v(E_{\mathrm{step}})$ is the velocity geometric factor \citep{2021ApJS..254...19M}, $g(t)$ is time-varying normalized SWAP efficiency, $\psi$ is the spacecrafts rotation angle, $ f(\vec{v})$ is the distribution function of the model populations (see Section \ref{subsec:distributions}), $v_{min}$ and $v_{max}$ are the limits of integration over speed $v$, calculated from the response function. We assume that the response function in the complementing angle ($R_{\phi}(\phi)$) is independent of energy and constant when the angle $\phi$ falls within the SWAP FOV, and zero otherwise.

\begin{figure*}[h]
    \centering
    \includegraphics[width=0.49\textwidth]{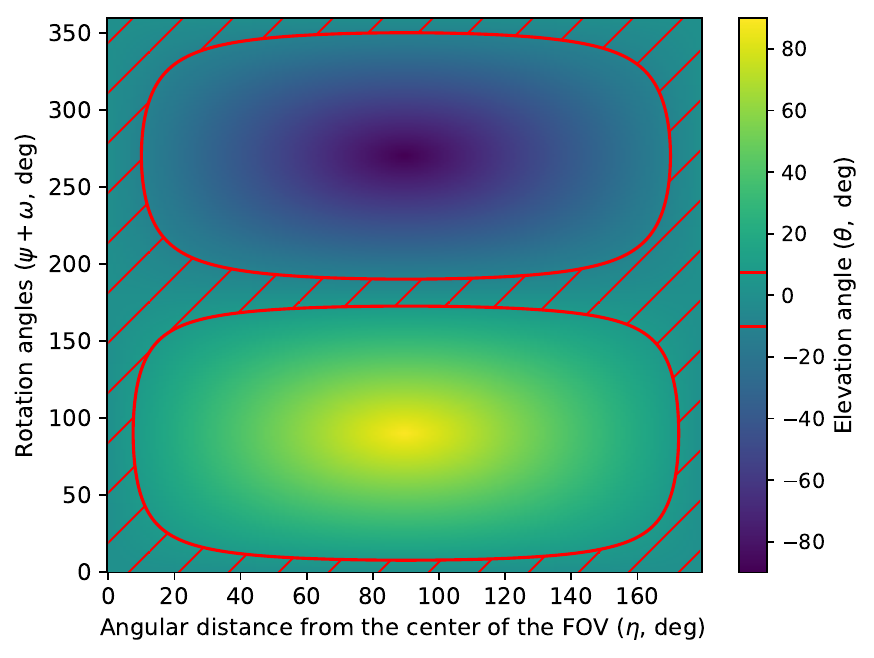}
    \includegraphics[width=0.49\textwidth]{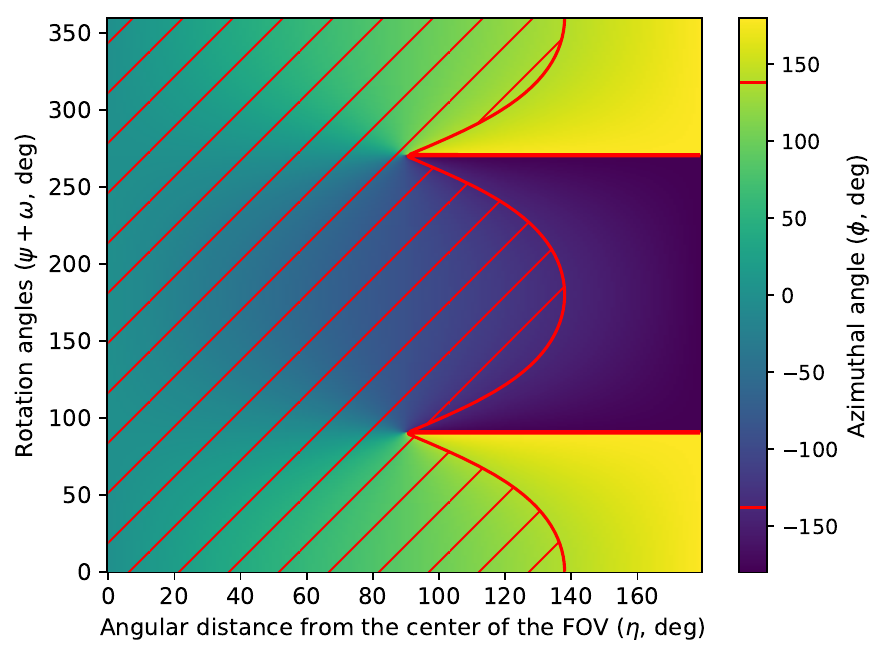}
    \caption{Mapping of the angular distance from the center of the FOV $\eta$ and rotation angles $\psi + \omega$ onto the elevation angle $\theta$ (\textit{left panel}) and azimuthal angle $\phi$ (\textit{right panel}). During a full rotation of the spacecraft around its axis, a given angular distance from the SWAP FOV center, $\eta$, corresponds to different values of the $\theta$ and $\phi$ coordinates, which can then be averaged. The SWAP FOV (hatched in red) limits the regions of the sphere that are observed by SWAP.}
    \label{fig:eta_theta_psi}
\end{figure*}

The novelty of our method is that we take into account the rotation of the spacecraft around its axis in the analysis of SWAP histogram data. Daily histograms consist of multiple measurements collected at various phases of the spacecraft rotation. Because energy stepping is not synchronized with this rotation, we average the integral over all spacecraft rotation angles (integral over $\psi$ in Equation (\ref{eq:int4})). To simplify the integral, we define new-rotating coordinates: the angular distance from the center of the FOV ($\eta$, [$0^\circ,180^\circ$]) and the rotational angle about the axis aligned with this center ($\omega$, [$0^\circ,360^\circ$]). Note that while the SWAP coordinates ($\theta$, $\phi$) rotate with the spacecraft, the new coordinates are in an inertial frame. The transitions from  the new coordinates to the SWAP FOV $\theta$ and $\phi$ coordinates are:
\begin{equation}\label{eq:thetaphi}
\begin{aligned}
    & \theta = \mathrm{arcsin}(\mathrm{sin} \, \eta \, \mathrm{sin}(\psi + \omega)), \\
    & \phi = \mathrm{arctan}(\mathrm{cos} \, \eta, \mathrm{sin} \, \eta \, \mathrm{cos}(\psi + \omega)).
\end{aligned}
\end{equation}
Figure \ref{fig:eta_theta_psi} shows the transitions from the newly defined angular distance from the center of the FOV $\eta$ and rotation angles $\psi + \omega$ to the elevation $\theta$ (left panel) and azimuthal $\phi$ (right panel) angles. Furthermore, we take into account that the center of the SWAP FOV is not pointed directly at the Sun, but at Earth for most of the observation time.

\begin{figure*}[h]
    \centering
    \includegraphics[width=0.49\textwidth]{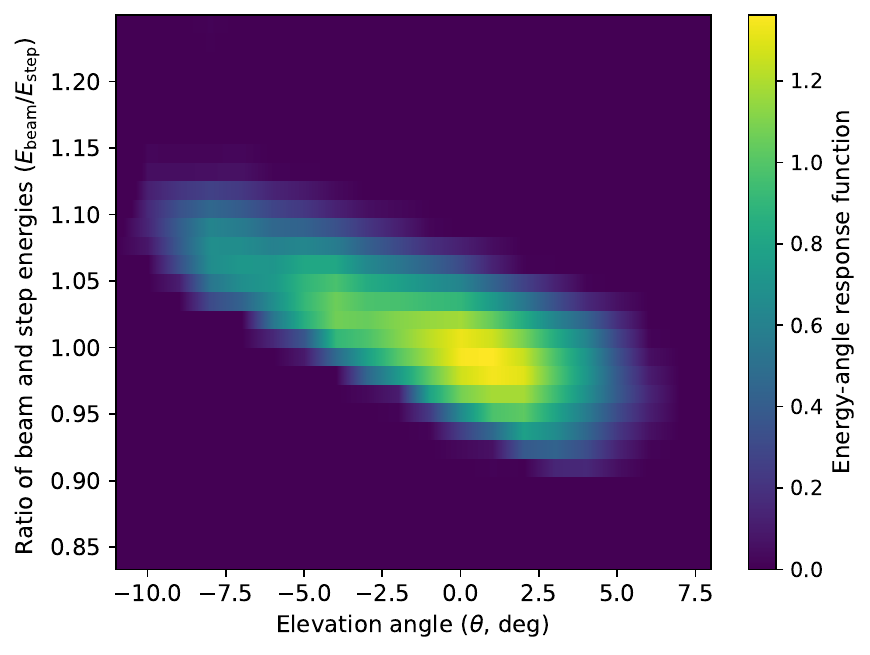}
    \includegraphics[width=0.49\textwidth]{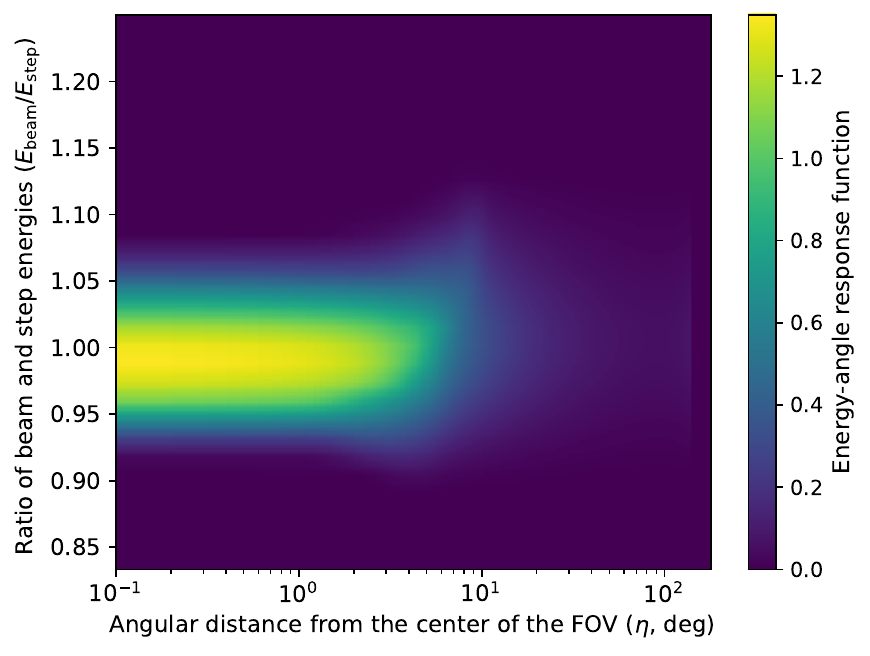}
    \caption{\textit{Left panel}: the averaged energy-angle response function $R_{\mathrm{E,\theta}}(E/E_{\mathrm{step}}, \theta)$ obtained based on data from \cite{elliott_new_2016}. \textit{Right panel}: the energy-angle response function $R_{\mathrm{E,\eta}}(E/E_{\mathrm{step}}, \eta)$ for rotation-average FOV used in our analysis. Using this new energy–angle response function simplifies the numerical calculations by eliminating one integration dimension. Instead of integrating over the SWAP $\theta$ and $\phi$ coordinates, it is sufficient to integrate over $\eta$.}
    \label{fig:response_function}
\end{figure*}

We transform the averaged energy-angle response function $R_{\mathrm{E,\theta}}(E/E_{\mathrm{step}}, \theta)$ (Fig. \ref{fig:response_function}, left panel) into a new energy-angle response function ($E_{\mathrm{beam}}/E_{\mathrm{step}}$ vs. $\eta$) (Fig. \ref{fig:response_function}, right panel) for rotation-average FOV: 
\begin{equation}\label{eq:R_int}
\begin{split}
    & R_{\mathrm{E,\eta}}(E/E_{\mathrm{step}}, \eta)  = \\ &\frac{1}{2\pi} \int_{0}^{2\pi}d\rho \, R_{\mathrm{E,\theta}}(E/E_{\mathrm{step}}, \mathrm{arcsin}(\mathrm{sin} \, \eta \, \mathrm{sin} \, \rho)) \, \\ & R_{\phi}(\mathrm{arctan}(\mathrm{cos} \, \eta, \mathrm{sin} \, \eta \, \mathrm{cos} \, \rho)),
    \end{split}
\end{equation}
where $\rho = \omega + \phi$.

Our new response function for rotation-average FOV is precalculated to simplify and speed up the numerical integration of the assumed solar wind and PUIs distributions. Instead of calculating the integral over four dimensions in Equation (\ref{eq:int4}), it is sufficient to calculate it over three dimensions: velocity $v$, distance from the center of the FOV $\eta$, and $\omega$ angle. The equation for model count numbers for a given energy step $C(E_{\mathrm{step}})$ is
\begin{equation}\label{eq:int3}
\begin{split}
    C(E_{\mathrm{step}}) = & \Delta t \, v_{\mathrm{c}} G^v(E_{\mathrm{step}}) g(t) \int_{0}^{2\pi}d\omega \int_{0}^{\eta_{max}} \mathrm{sin} \, \eta \, d\eta  \\& \int_{v_{min}}^{v_{max}}dv \, f(\vec{v}) R_{\mathrm{E,\eta}}(E/E_{\mathrm{step}}, \eta) v^3.
    \end{split}
\end{equation}

We select the nodes at which the energy-angle response function is calculated to optimize the integration of models of the solar wind and PUIs. In particular, a denser grid at small values of $\eta$ is required for accurate integration of the kappa distribution for the core solar wind. For PUIs, a uniform but sufficiently dense grid of energy-angle response values is needed. We test several numerical integration methods. Gauss–Legendre integration proves to be the most effective for the core solar wind. Additionally, we examine whether integration over the entire SWAP FOV is necessary. For both protons and alpha particles, this is not required.  Depending on the solar wind temperature, 99\% of the signal is collected from an angular range spanning only a few to eighty degrees. Therefore, we define different angular limits ($\eta_{\mathrm{max}}$) and integration grids separately for hydrogen and helium. For PUIs, the distribution is more uniform, and integration using the trapezoidal rule is sufficient. However, the full-field integration over the entire SWAP FOV is necessary in this case.

\subsection{Distribution functions of solar wind ions and PUIs} \label{subsec:distributions}

We assume that the core solar wind populations (protons, alpha particles, $\mathrm{He^+}$) follow the kappa distribution \citep{livadiotis_first_2011,  randol_observations_2012, nicolaou_properties_2014,2017ApJS..233....8M,swaczyna_non-equilibrium_2019, 2021ApJS..254...19M, 2022ApJ...934..147M}, given as:
\begin{equation}
\begin{split}
f(\bm{v}; n, \theta_{\mathrm{T}}, \kappa ) = &
\frac{n}{(\pi \theta_{\mathrm{T}}^2)^{3/2} (\kappa - 3/2)^{3/2}} 
\frac{\Gamma(\kappa+1)}{\Gamma\!\left(\kappa - 1/2\right)} \\ &
\left( 1 + \frac{1}{\kappa - 3/2} \frac{(\bm{v}-\bm{u})^2}{\theta_{\mathrm{T}}^2} \right)^{-\kappa - 1},
\end{split}
\end{equation}
where $\bm{v}$ is the velocity vector, $n$ is the density, $\bm{u}$ is the bulk solar wind velocity, $\theta_{\mathrm{T}}$ is the effective speed scale parameter, and $\kappa$ is the kappa index. The speed scale parameter is defined as $\theta_{\mathrm{T}} = \sqrt{\frac{2 k_{\mathrm{B}} T}{m}}$ where $k_{\mathrm{B}}$ is the Boltzmann constant, $T$ is the ion temperature, and $m$ is the mass of the particle.

In contrast, the energy spectrum of PUIs differs significantly from that of the core solar wind ions \citep{1976JGR....81.1247V}. We assume that PUIs are represented by the generalized filled shell model \citep{2014JGRA..119.7142C, 2015JGRA..120.9269C}, which was used in the previous New Horizons/SWAP studies \citep{swaczyna_density_2020,2022ApJ...934..147M}:
\begin{equation}\label{eq:PUI_Chen}
\begin{split}
f(\bm{r}, w) = & \frac{1}{4\pi} \frac{\beta_0 r_0^2}{u_{\mathrm{sw}}} \frac{\alpha S(\bm{r}, w)}{r v_{\mathrm{b}}^3} w^{\alpha-3} n_{\mathrm{H,TS}} \\ & \exp{ \left(-\frac{\lambda}{r} \frac{\theta}{\sin{\theta}} w^{-\alpha} \right)} \Theta \left(1-w\right),
\end{split}
\end{equation}
where $\bm{r}$ is the heliocentric position, $w = v/v_{\mathrm{b}}$ is the ratio of the PUI speed $v$ to the injection speed $v_\mathrm{b}$, $\beta_{\mathrm{0}}$ is the ionization rate normalized to $r_0$ = 1 au, $u_{\mathrm{sw}}$ is the bulk speed of the solar wind in the solar frame, $\alpha$ is the cooling index, and $S(\bm{r}, w)$ is the survival probability of PUIs from their production distance to the point of observation \citep[Eq. 3]{swaczyna_density_2020}, $n_{\mathrm{H,TS}}$ is the density of the ISN at the upwind termination shock, $\lambda$ in the ISN ionization cavity size (4 au for the ISN hydrogen and 0.5 au for the ISN helium \citep{2021ApJS..254...19M, swaczyna_interstellar_2024}), $\theta$ is the angle between radial and the ISN inflow direction, and $\Theta$ is the Heaviside step function.

\subsection{Fitting method and parameters} \label{subsec:models}

Unlike the previous studies that relied on the $\chi^2$ fitting method, we minimize the maximum likelihood estimator to find the best-fit parameters for each population. The maximum likelihood method remains statistically valid even when the number of counts is small. It is based on comparing the counts in each energy-per-charge bin predicted by the used model ($C_\mathrm{mod}$) with those observed ($C_\mathrm{obs}$). The minimized likelihood function ($L$) is given as \citep[e.g.,][]{1984NIMPR.221..437B}:
\begin{equation}\label{eq:MLE}
    L = -2\sum_{i} (C_{\mathrm{obs},i} - C_{\mathrm{mod},i} +  C_{\mathrm{obs},i}\mathrm{log}\frac{C_{\mathrm{mod},i}}{C_{\mathrm{obs},i}}).
\end{equation}
To calculate predicted counts for a given energy step, the distribution function is integrated as shown in Equation (\ref{eq:int3}), and the spacecraft velocity is taken into account.

\begin{figure*}[h]
    \centering
    \includegraphics[width=\linewidth]{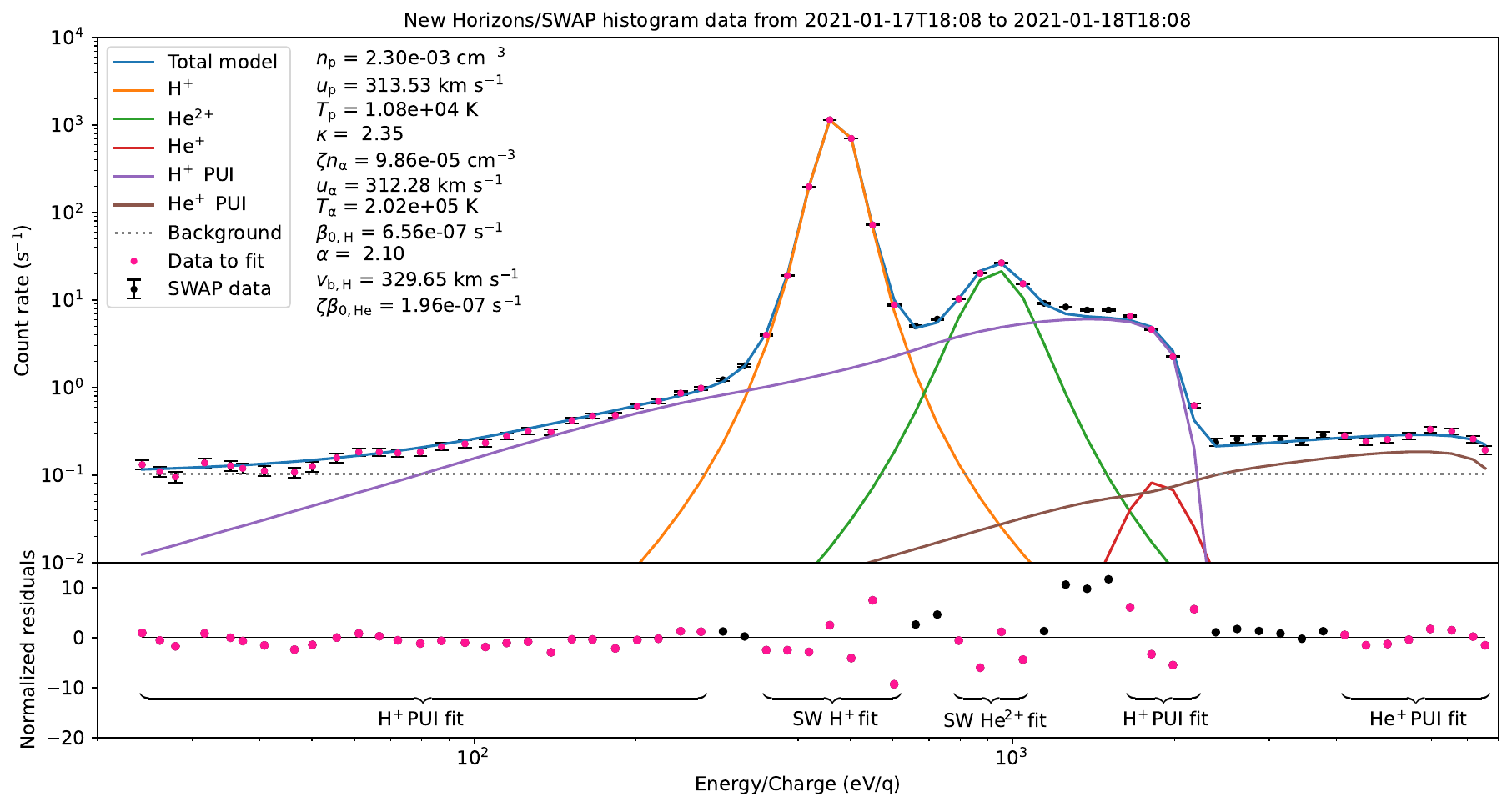}
    \caption{\textit{Upper panel:} Example of a SWAP daily averaged spectrum with Poisson uncertainties (black dots with error bars) with fit models (solid lines). The graph shows both the core solar wind ($\mathrm{H^+}$, $\mathrm{He^{2+}}$, $\mathrm{He^+}$) and PUIs ($\mathrm{H^+}$ PUI, $\mathrm{He^+}$ PUI). The core solar wind protons are shown in orange, alpha particles in green, $\mathrm{He^+}$ ions in red, $\mathrm{H^+}$ PUIs in purple, and $\mathrm{He^+}$ PUIs in brown. The data points used for the fit are marked in pink (cf. text). The gray dashed line is the background. The fit values of the most important parameters are listed in the figure. \textit{Lower panel:} Residual signal between the total model and SWAP data normalized by the statistical uncertainty, with a thin black line marking zero. In the selected case, the low solar wind speed shifts the energy cutoff of the $\mathrm{He^+}$ PUI distribution close to the upper limit of the SWAP energy range.}
    \label{fig:fit_example}
\end{figure*}

We individually fit the appropriate models to each of the daily spectra to obtain the physical parameters of the solar wind and PUIs (Fig. \ref{fig:fit_example}). For core solar wind protons (orange line), we fit four parameters ($n_{\mathrm{p}}$, $u_{\mathrm{p}}$, $T_{\mathrm{p}}$, $\kappa$). For core solar wind alpha particle (green line), we assume the same $\kappa$ parameter as for protons, and fit the other three parameters ($n_{\mathrm{\alpha}}$, $u_{\mathrm{\alpha}}$, $T_{\mathrm{\alpha}}$, $n_{\mathrm{\alpha}}$). In the fitting procedure, we assume the same detection efficiency for protons and helium ions. Nevertheless, we estimate the relative helium-to-hydrogen detection efficiency (denoted as $\zeta$) in Section \ref{subsec:results_efficiency}. To indicate that the fitted alpha particle density is scaled by this factor, we report it as $\zeta n_{\mathrm{\alpha}}$. Due to their minor contribution to the overall spectrum, we do not fit the parameters for solar wind $\mathrm{He^+}$ (red line), but we assume that their velocity, temperature, and kappa are the same as those of alpha particles, and estimate the density of solar wind $\mathrm{He^+}$ ($n_{\mathrm{He^+}}$) based on the cumulative probability of production of an $\mathrm{He^+}$ ion from a solar wind alpha particle ion on its path from the Sun to the distance $r$ \citep[Eq. 6]{swaczyna_he_2019}.

We fit three parameters defining the distribution function of hydrogen PUIs (Fig. \ref{fig:fit_example}, purple line): $\beta_{\mathrm{0,H}}$, $\alpha$, and $v_{\mathrm{b,H}}$. In the context of the entire study, the most important aspect is the distribution of helium PUIs (Fig. \ref{fig:fit_example}, brown line). We assume the same cooling index as for $\mathrm{H^+}$ PUIs, and estimate the $\mathrm{He^+}$ PUI injection speed ($v_{\mathrm{b,He}}$) from:
\begin{equation}\label{eq:hep}
    v_{\mathrm{b,He}} = v_{\mathrm{b,H}}\frac{|\textbf{v}_{\mathrm{He}}-\textbf{u}_{\mathrm{p}}|}{|\textbf{v}_{\mathrm{H}}-\textbf{u}_{\mathrm{p}}|},
\end{equation}
where $\textbf{v}_{\mathrm{H}}$ is the ISN hydrogen velocity \citep[22 $\mathrm{km \, s^{-1}}$,][]{2005Sci...307.1447L},  $\textbf{v}_{\mathrm{He}}$ is the ISN helium velocity \citep[25.4 $\mathrm{km \, s^{-1}}$,][]{2015ApJS..220...28B, 2015ApJS..220...22M}. Finally, we fit the helium ionization rate normalized to 1 au ($\beta_{\mathrm{0,He}}$), which is the key variable. This ionization rate is also scaled by the relative helium-to-hydrogen detection efficiency ($\zeta$). All fit parameters are summarized in Table \ref{tab:parameters_names}.

\begin{deluxetable}{lll}
\tablecaption{Fit parameters for solar wind and PUI populations. \label{tab:parameters_names}}
\tablehead{
\colhead{Population} & \colhead{Parameter} & \colhead{Symbol}
}
\startdata                                                                          
{} &  proton density & $n_{\mathrm{p}}$ \\  
H$^{+}$   & proton bulk speed & $u_{\mathrm{p}}$ \\
{} & proton temperature  & $T_{\mathrm{p}}$ \\
{} & kappa index &  $\kappa$ \\ \hline
{} &  alpha density & $n_{\mathrm{\alpha}}$ \\  
He$^{2+}$   & alpha bulk speed & $u_{\mathrm{\alpha}}$ \\
{} & alpha temperature  & $T_{\mathrm{\alpha}}$ \\ \hline
{} & hydrogen ionization rate normalized to 1 au   &   $\beta_{\mathrm{0,H}}$  \\ 
H$^{+}$   PUI & cooling index &  $\alpha$    \\ 
{}  & hydrogen PUI injection speed   &   $v_{\mathrm{b,H}}$    \\ \hline
He$^{+}$   PUI  & helium ionization rate normalized to 1 au   &  $\beta_{\mathrm{0,He}}$
\enddata
\end{deluxetable}

\subsection{Selecting data points for model fits}
\label{subsec:selection}

An important aspect of our method is the selection of appropriate data points used to fit the individual components of the SWAP histogram-type spectrum (indicated by pink dots in Fig. \ref{fig:fit_example}). We execute the entire fit procedure in two iterations, and the selected data points could differ between them. The algorithm we developed is described below, and the summary is presented in Table \ref{tab:parameters}.

We start by estimating the background (dotted gray line) based on the average count rate in the first four energy bins. In this energy range, we do not expect a significant contribution from the core solar wind, only a minor contribution from $\mathrm{H^+}$ PUIs. The background is then added to all model spectra. Subsequently, based on the location of the maximum of the entire spectrum, we identify the core solar wind hydrogen ions. In the first iteration, the energy bin with the highest number of counts, along with the four adjacent bins (two on each side), is selected to fit the solar wind proton distribution. Next, we calculate the energy bin corresponding to the maximum of the alpha particle distribution. The calculated bin is typically a local maximum of the spectrum. To fit the alpha particle distribution, we take this bin, the bin next to it that is larger, and two bins adjacent to these two. Including only the bins with the highest count rates in the fit of the proton and alpha models minimizes the contribution from $\mathrm{H^+}$ PUIs. We then determine the core solar wind $\mathrm{He^+}$ model based on the density determined from alpha particle partial neutralization \citep[Eq. 6]{swaczyna_he_2019}.

Next, we select data points to fit the $\mathrm{H^+}$ PUIs component. At low energies, all energy bins up to two bins before the lowest-energy bin selected to fit the protons are initially considered. If, at any point, the sum of the core solar wind proton and alpha particle models exceeded 10\% of the observed counts in a given bin, that bin is excluded from the fit. This approach ensures that the fit includes only data where the contribution of the modeled PUIs is genuinely dominant. To accurately fit the $\mathrm{H^+}$ PUIs population, we include observations near the theoretical cut-off energy. We estimate the cut-off energy based on the solar wind bulk speed and the speed of ISN hydrogen. Four bins with the energies lower than the estimated cut-off energy are selected, as well as additional bins with higher energies, provided that their counts exceeded 10\% of the maximum count among the already selected points. These additional bins are added to correctly capture the cut-off energy if the injection speed is much higher than the calculated one. The number of additional points varies from day to day. In most cases, they are not needed, but occasionally a few must be added above the initially estimated theoretical energy cutoff. In the first iteration, we add the previously calculated core solar wind model to the $\mathrm{H^+}$ PUIs model.

The final step is the selection of data points to fit the $\mathrm{He^+}$ PUIs model. The cutoff of its spectrum is not visible because the SWAP energy range is not wide enough. However, we include all energy bins with energies greater than eight times the energy corresponding to the maximum of the proton distribution. If no such bins exist, only one last energy bin is used. However, in the helium ionization rates analysis (see Section \ref{subsec:beta}), we consider only those observation days that had at least two energy bins exceeding eight times the energy corresponding to the proton peak. This selection criterion is applied to ensure reliable ionization rate estimates. By taking into account only the highest energy bins, we minimize the likelihood of spectrum contamination by unmodeled elements, such as carbon, oxygen, and iron. During coronal mass ejections (CME), the densities of $\mathrm{C^{2+}}$ and $\mathrm{O^{2+}}$  are even on the order of $10^{-5} \, \mathrm{cm}^{-3}$ at 1 au \citep{gilbert_first_2012}, so in our method we exclude energy bins where their contribution could be significant. Only $\mathrm{C^+}$, $\mathrm{O^+}$, and $\mathrm{Fe^{4+}}$, $\mathrm{Fe^{5+}}$, and $\mathrm{Fe^{6+}}$ fall within the SWAP energy range used for the $\mathrm{He^+}$ PUIs fit, but their contributions to the overall spectrum are negligible \citep{2010A&A...512A..72G}. Even during CME, when lower ionic charge states are expected, the listed carbon, oxygen, and iron ions should not affect the fit of the $\mathrm{He^+}$ PUIs model \citep{gilbert_first_2012}.

The second fit iteration follows the same rules as described above, with a few modifications. The most significant modification is the inclusion of the $\mathrm{H^+}$ PUIs distribution obtained in the first iteration. We subtract it when calculating the background from the average count rate in the first four energy bins. We add the $\mathrm{H^+}$ PUIs distribution to the estimated counts in the proton and alpha models. Moreover, for fitting the solar wind $\mathrm{H^+}$ ions, seven data points are selected instead of five, with one bin added at both the lower and upper ends of the original interval. Additionally, when fitting the $\mathrm{H^+}$ PUI model, we add the previously calculated counts from the core solar wind and the $\mathrm{He^+}$ PUIs.

\begin{deluxetable*}{llll}
\tablecaption{Parameters and number of data points used for each fit for a given population \label{tab:parameters}}
\tablehead{
\colhead{Population} & \colhead{No of fit parameters} & \colhead{Iteration 1: No of data points} & \colhead{Iteration 2: No of data points}
}
\startdata
H$^+$         & 4 ($n_{\mathrm{p}}$, $u_{\mathrm{p}}$, $T_{\mathrm{p}}$, $\kappa$)                      & 5                                  & 7              \\ \hline
He$^{2+}$        & 3 ($n_{\mathrm{\alpha}}$, $u_{\mathrm{\alpha}}$, $T_{\mathrm{\alpha}}$)                        & 4                                  & 4              \\ \hline
H$^+$  PUI     & 3 ($\beta_{\mathrm{0,H}}$, $\alpha$, $v_{\mathrm{b,H}}$)                        & $\sim$30                           & $\sim$30       \\ \hline
He$^+$  PUI    & 1 ($\beta_{\mathrm{b,He}}$)       & $\geq$2 & $\geq$2
\enddata
\end{deluxetable*}

\section{Results} \label{sec:results}

Fitting models to SWAP observational data allows us to determine the physical parameters of the core solar wind and PUIs. For protons, alpha particles, and $\mathrm{He^+}$, the fit parameters are the density, speed, temperature, and the kappa parameter. For $\mathrm{H^+}$ and $\mathrm{He^+}$ PUIs, the fit parameters are the ionization rate, cooling index, and PUI injection speed. The estimated parameters of the core solar wind and PUIs are used in the further stages of the analysis. The fit parameters are available in the data file supplied with this article.

However, the determined densities of helium ions and the $\mathrm{He^+}$ PUIs ionization rate are scaled by the detection efficiency for helium ions compared to protons (denoted as $\zeta$). The value of this parameter for the SWAP instrument is not precisely known, although in previous works it was assumed to be 1.5 \citep[e.g.,][]{swaczyna_he_2019}. In order to correctly interpret the helium ionization rates, we aim to find a better estimate for this parameter by comparing the solar wind composition observed by SWAP with 1 au observations.

\subsection{SWAP efficiency for helium} \label{subsec:results_efficiency}

To estimate the efficiency of the SWAP instrument for helium relative to hydrogen (the $\zeta$ parameter), we use the results of our analysis and data from the OMNI 2 database \citep{2005JGRA..110.2104K}. This database provides solar wind magnetic field and plasma parameters collected since 1963 by various spacecraft in geocentric or L1 (Lagrange point) orbits. For our analysis, we extract the daily-averaged data on the alpha to proton density ratio. We use the alpha to proton density ratio to determine the $\zeta$ parameter, because as the plasma propagates away from the Sun, both alpha and proton densities decrease due to expansion, but their ratio should remain constant. During the considered period, the main OMNI data source is the Wind spacecraft.

The alpha to proton density ratio $(n_{\mathrm{\alpha}}/n_{\mathrm{p}})_{\mathrm{OMNI}}$ from OMNI is measured near Earth. Since the SWAP observations are collected in the outer heliosphere, far from 1 au, some alpha particles are partially decharged to $\mathrm{He^+}$ and some protons are neutralized. Therefore, to compare the compositions observed of SWAP and OMNI, we need to adjust the SWAP-observed values to compensate for these processes. We add the core solar wind $\mathrm{He^+}$ density, $n_{\mathrm{He^+}}$, to the alpha particle density, $n_{\mathrm{\alpha}}$. We modify the proton density, $n_{\mathrm{p}}$, by adding the fraction of PUIs created in charge exchange collisions from protons \citep[Eq. 13]{swaczyna_density_2020}. 

We time-shift the SWAP alpha to proton density ratio corrected for the above effects, $(\zeta n_{\mathrm{\alpha}}/n_{\mathrm{p}})_{\mathrm{SWAP, corr}}$, to correspond to measurements at 1 au. For each daily SWAP spectrum, we estimate the bulk speed ($u_{p}$), proton density ($n_p$), and the $\mathrm{H^+}$ PUIs ionization rate ($\beta_{\mathrm{0,H}}$). Based on these parameters, we calculate the $\mathrm{H^+}$ PUIs density ($n_{\mathrm{H \, PUI}}$) \citep[Eq. 10]{swaczyna_density_2020}. We obtain the average solar wind speed \citep[Eq. 12]{swaczyna_density_2020} for each day, taking into account the slowing of the solar wind due to the PUI production. We divide all the data into 27-day Carrington rotation periods and average the solar wind parameters over them. For every period with 20 valid SWAP observation days, we calculate the time required for the solar wind to travel from 1 au to the New Horizons location ($r$). In this way, we obtain time-shifted alpha to proton density ratios averaged over Carrington periods, determined from SWAP $(\zeta n_{\mathrm{\alpha}}/n_{\mathrm{p}})_{\mathrm{SWAP, av}}$. Finally, we average the $(n_{\mathrm{\alpha}}/n_{\mathrm{p}})_{\mathrm{OMNI, av}}$ ratio from OMNI data over a time window of $\pm$12.5 days around the middle of the Carrington period for the time-shifted date. 

\begin{figure*}[h]
    \centering
    \includegraphics[width=0.49\linewidth]{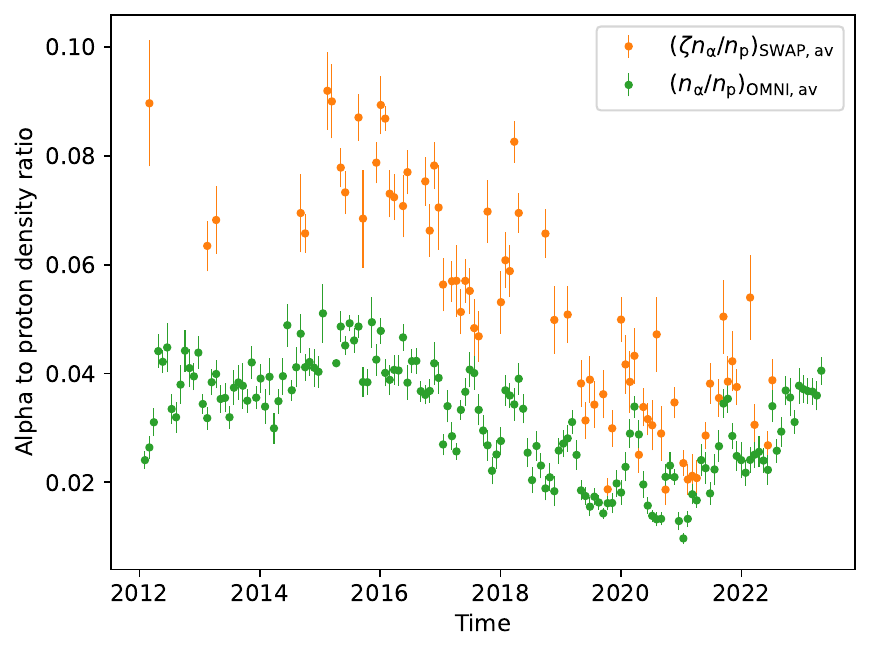}
    \includegraphics[width=0.49\linewidth]{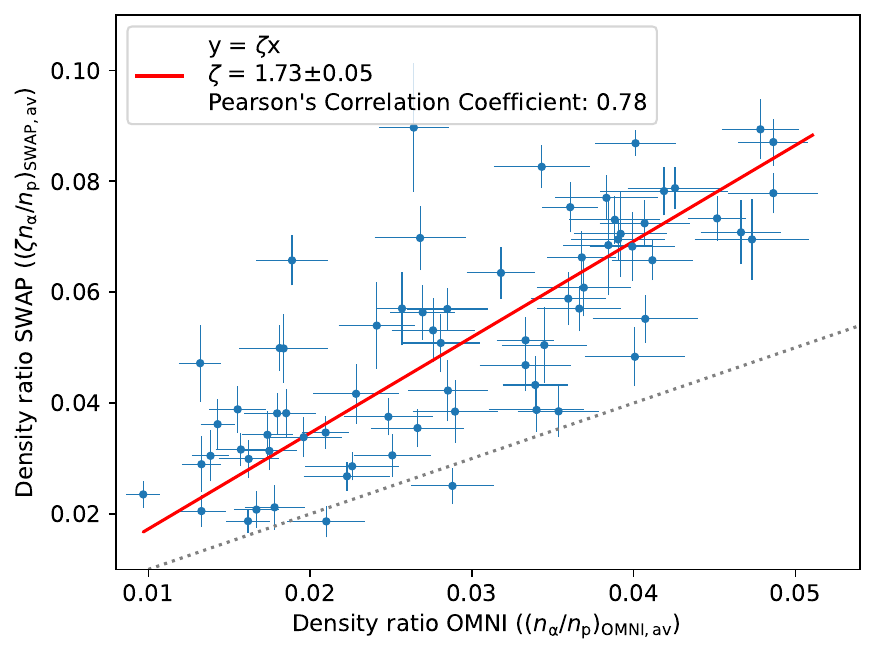}
    \caption{\textit{Left panel:} The time variations of the alpha to proton density ratios from SWAP (orange) and from OMNI 2 database (green). The alpha to proton density ratios determined from the SWAP data are larger than those obtained from the OMNI 2 database by the $\zeta$ scaling factor. \textit{Right panel:} The relation between the alpha to proton density ratio from SWAP and OMNI 2 database. The gray dashed line represents $\zeta$ = 1. The red line shows the fit of the scaling factor $\zeta$.}
    \label{fig:scaling_parameter}
\end{figure*}

The left panel of Figure \ref{fig:scaling_parameter} shows time variations of the alpha to proton density ratios from SWAP $(\zeta n_{\mathrm{\alpha}}/n_{\mathrm{p}})_{\mathrm{SWAP, av}}$ and from the OMNI 2 database $(n_{\mathrm{\alpha}}/n_{\mathrm{p}})_{\mathrm{OMNI, av}}$. The right panel of Figure \ref{fig:scaling_parameter} shows the relation between the alpha to proton density ratio from SWAP and OMNI. We assume that the values should be proportional to each other due to the differences in the SWAP instrument efficiency for hydrogen and helium ($\zeta$ parameter). The estimated helium-to-hydrogen detection efficiency ratio is $1.73 \pm 0.05$. This value is greater than the generally assumed difference in the previously used SWAP efficiency for helium relative to hydrogen of 1.5 \citep[e.g.,][]{swaczyna_he_2019}. The Pearson correlation coefficient of approximately 0.78 indicates a strong correlation between the SWAP and 1 au measurements. Note that the plasma parcels observed by SWAP are not the same as observed at 1 au. Therefore, this correlation is not expected to be perfect.

\subsection{Helium ionization rate} \label{subsec:beta}

The ionization rates obtained by us from the $\mathrm{He^+}$ PUIs observations are also scaled by the relative helium-to-hydrogen efficiency. Therefore, we divide the fit ionization rates by the scaling factor of 1.73 found in Section \ref{subsec:results_efficiency}. 

\begin{figure*}[h]
        \centering
        \includegraphics[width=0.49\linewidth]{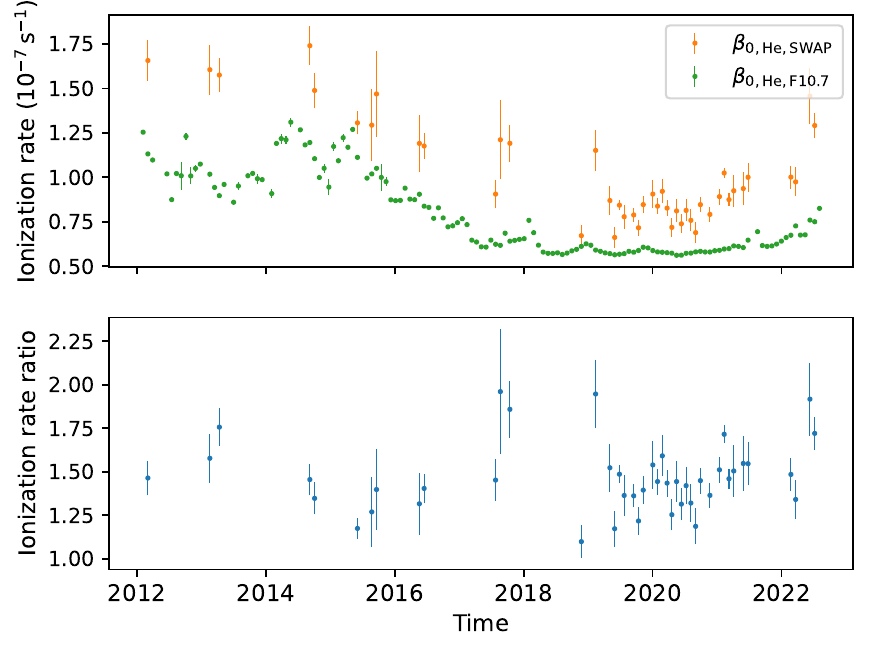}
        \includegraphics[width=0.49\linewidth]{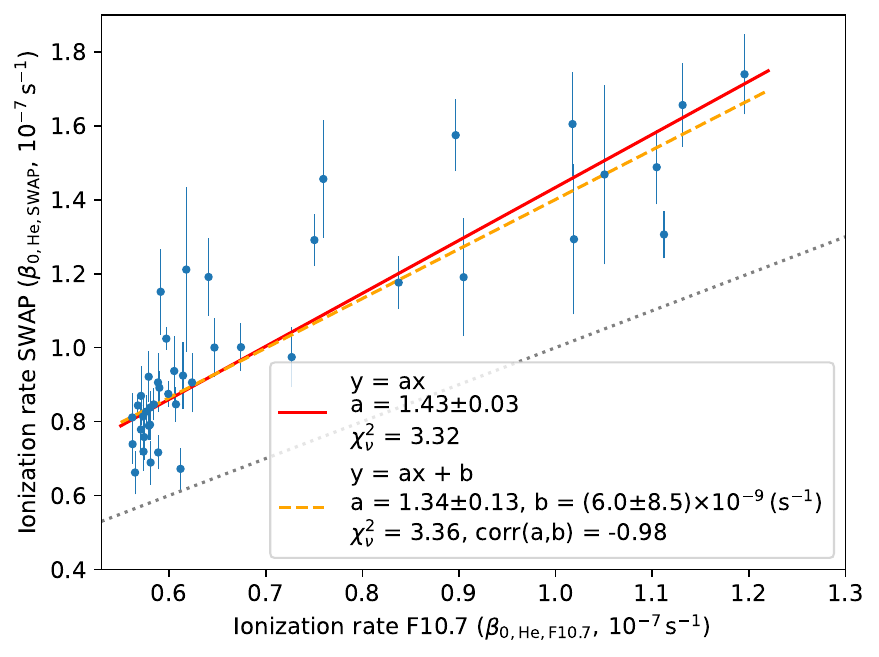}
        \caption{\textit{Left panel:} The time variations of the helium SWAP ionization rates (orange) and the photoionization rates derived from the model (green), and their ratio. The helium ionization rate ratio indicates that the SWAP ionization rates are higher than the photoionization rates from solar UV spectra and 10.7 radio flux. \textit{Right panel:} The relation between helium SWAP ionization rates and photoionization rates with uncertainties. The gray dashed line represents a one-to-one relationship. The red solid line shows the model assuming proportionality of values, and the orange dashed line is the linear model. The model with only the scaling parameter fit shows that the helium ionization rates estimated from SWAP observations are 43\% higher than the photoionization rates based on solar UV spectra and 10.7 radio flux.}
        \label{fig:beta}
\end{figure*}

We compare the derived ionization rates with the photoionization rates of helium ($\beta_{\mathrm{0, He, F10.7}}$) calculated based on a series of observations of the solar EUV spectrum by TIMED correlated with the solar radio flux in 10.7 cm \citep{, 2018SoPh..293...76W, 2020ApJ...897..179S}. Photoionization is the dominant ionization process of ISN helium, especially at distances greater than a few astronomical units from the Sun. The solar wind takes several months to travel to the location of New Horizons. Therefore, before comparing the photoionization rates with the SWAP measurements for each observation day, the data have to be properly averaged. We average the photoionization rates along the expanding solar wind from 1 au to New Horizons. The time variations of the SWAP ionization rates $\beta_{\mathrm{0, He, SWAP}}$ and averaged photoionization rates $\beta_{\mathrm{0, He, F10.7}}$, and their ratio $\frac{\beta_{\mathrm{0, He, SWAP}}} {\beta_{\mathrm{0, He, F10.7}}}$ is shown in Figure \ref{fig:beta}, left panel.

We also divide the entire observation period into intervals of one solar rotation. Both the SWAP-derived ionization rates and the photoionization rates are averaged over these intervals. A comparison of the results is shown in Figure \ref{fig:beta}, right panel. All data points lie above the gray line representing the equal ionization rates. We fit two models to the data. The first assumes proportionality of the data (red solid line) and the second uses general linear dependence (orange dashed line). Due to the strong correlation between parameters $a$ and $b$ in the linear fit (-0.98) and similar $\chi^2$ values, we conclude that a model with only the scaling parameter fit is sufficient. Therefore, the helium ionization rates estimated from SWAP observations are an average $43\%$ higher than the photoionization rates based on a solar UV spectra and 10.7 radio flux.

\section{Discussion} \label{sec:discussion}

\subsection{Uncertainty of the relative helium-to-hydrogen SWAP efficiency} \label{subsec:discussion_density}

The estimated SWAP efficiency for helium relative to hydrogen is determined based on the entire dataset available after the initial selection (see Section \ref{subsec:results_efficiency}). We find that the $\zeta$ scaling factor is $1.73 \pm 0.05$. However, the reported uncertainty represents only the statistical uncertainty, while further systematic sources also affect this factor.

We try to estimate systematic uncertainties by splitting the dataset with respect to the solar wind parameters. We divide the dataset into two subsets, first by the median proton speed and then by the proton temperature. In both cases, noticeable differences are observed. For the slower solar wind, the scaling factor is $1.65 \pm 0.08$, while for the faster solar wind it is $1.78 \pm 0.06$. Similarly, for the cooler solar wind, the factor is $1.71 \pm 0.08$, and for the hotter solar wind, $1.75 \pm 0.06$. The obtained spread between these cases provides a measure of the uncertainty of this factor. The results are statistically consistent with each other.

\subsection{Time variation of ISN helium density} \label{subsec:discussion_ionization}

The $\mathrm{He^+}$ PUIs model is proportional to both the ISN helium density ($n_{\mathrm{ISN \, He}}$) and the ionization rate ($\beta_{\mathrm{0, He}}$).  Therefore, in the fitting, we cannot determine these two factors separately. If we take a constant ISN helium density of 0.015 cm$^{-3}$  \citep{2004A&A...426..845G}, the average helium ionization rates ($\beta_{\mathrm{0, He, SWAP}}$) derived from SWAP helium PUIs observations are approximately 43\% higher than the photoionization rates ($\beta_{\mathrm{0, He, F10.7}}$) calculated based on a series of solar spectra observed by TIMED correlated with the solar 10.7 flux. (see Section \ref{subsec:beta}). However, if we assume the photoionization rates from the F10.7 flux, we can attribute the changes to the time variation of the ISN helium density. This effect might be responsible for the modulation of IBEX-Lo data previously described by \cite{swaczyna_very_2022} based on IBEX observations from 2009 to 2020. Contrary to the $\mathrm{He^+}$ PUIs observations from SWAP, increasing the ionization rate causes a decrease in the flux of ISN helium reaching IBEX-Lo at 1 au. 

We attempt to investigate whether such a variability in the ISN helium density from New Horizons may explain both observations. The SWAP data analyzed in this study span the years 2012 to 2022. Assuming that ISN helium atoms propagate from the VLISM toward the Sun, ionized helium atoms are first observed as PUIs by the SWAP and subsequently detected by the IBEX-Lo instrument at 1 au. Considering that the ionization cavity can be neglected compared to the New Horizons distance,  $\mathrm{He^+}$ PUIs observed by SWAP are on average formed roughly mid-way between the Sun and New Horizons. Therefore, to compare IBEX-Lo and SWAP observations, we time-shift the observations for the time needed for ISN helium atoms to travel from this distance to 1 au. The period covered in \cite{swaczyna_very_2022} overlaps with the SWAP observation window through 2017.

\begin{figure}[h]
    \centering
    \includegraphics[width=\linewidth]{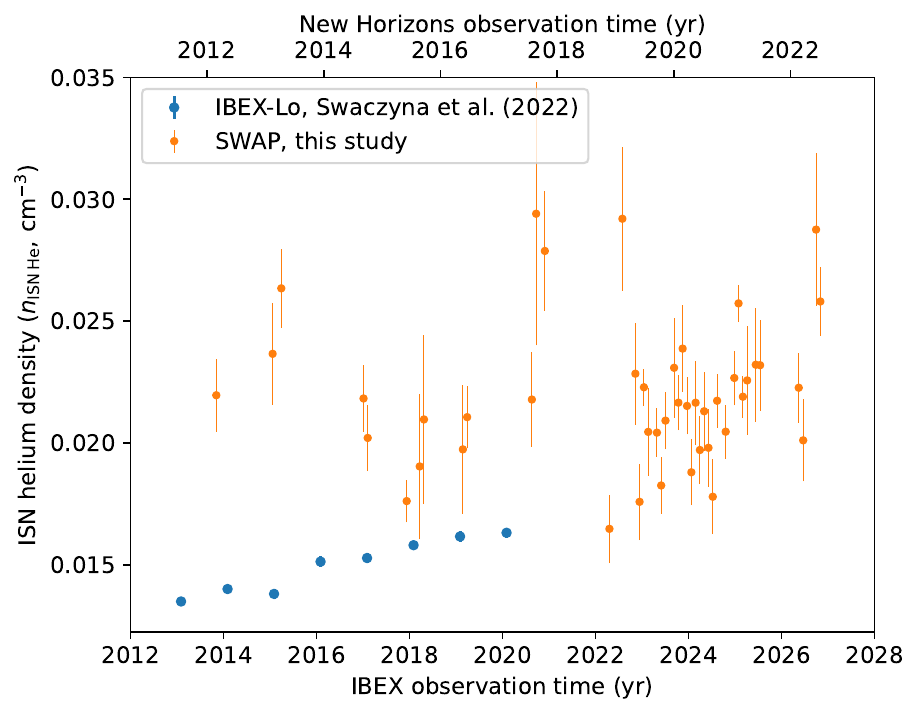}
    \caption{Estimated ISN helium density from SWAP (orange dots) and IBEX-Lo (blue dots) data with uncertainties. The values from IBEX-Lo are calculated by normalizing linear coefficients from Figure 1 in \cite{swaczyna_very_2022} and multiplying them by the ISN helium density of 0.015 cm$^{-3}$. The ISN helium densities inferred from SWAP and IBEX-Lo observations show opposite trends over the overlapping periods.}
    \label{fig:IBEX}
\end{figure}

To test the hypothesis of temporal variability in ISN helium density, we estimate its values based on observations from both the IBEX-Lo and SWAP instruments. For IBEX-Lo, we use the normalized linear coefficients from Figure 1 in \cite{swaczyna_very_2022} and multiply them by an assumed constant ISN helium density of 0.015 cm$^{-3}$, to reconstruct its variation over time (Fig. \ref{fig:IBEX}, blue dots). For SWAP, we compute an analogous estimate by multiplying the same assumed constant helium density by the ratio of the $\beta_{\mathrm{0, He, SWAP}}$ to $\beta_{\mathrm{0, He, F10.7}}$ Fig. \ref{fig:IBEX}, orange dots).

During the period when both instruments provide coverage, SWAP indicates a decrease in the density, whereas IBEX suggests an increase. Such a change cannot be explained by the uncertainty of the geometric factor. Therefore, the hypothesis of a time-varying ISN helium density appears to be incorrect. Unfortunately, the New Horizons dataset is incomplete in the overlapping period. However, the more complete observations were collected by SWAP between 2019 and 2021. Those data can be compared with 1 au observations of ISN helium between 2023 and 2026.

Given the unlikely nature of a time-varying ISN helium density, we conclude that the helium ionization rate is approximately 43\% higher than that predicted by models. Even accounting for a maximum 6\% contribution to the total ionization from charge exchange \citep{2019ApJ...872...57S}, the discrepancy between the models and our result remains significant. It is unlikely that the obtained ionization rate is overestimated due to electron impact ionization, which decreases faster than the inverse square of the distance from the Sun, as it strongly depends on the temperature of electrons that rapidly drops in the outer heliosphere \citep{1994JGR....9923401S, 1998JGR...103.1969I, 2000JGR...10518337M}.

\section{Summary} \label{sec:summary}

In this paper, we use $\mathrm{He^+}$ PUI observations from the SWAP instrument onboard New Horizons to determine the temporal evolution of helium ionization rates. We use the SWAP observations from 2012-2022 corresponding to distances of 22-54 au from the Sun. We fit models of the solar wind protons, alpha particles, as well as $\mathrm{H^+}$ and $\mathrm{He^+}$ PUIs to energy-per-charge daily spectra from SWAP. This paper presents a new method that accounts for the SWAP energy-angle response \citep{elliott_new_2016} in integrating the model distribution functions with the instrument response. The method utilizes the fact that the histogram data used here consist of multiple scans performed at various spin phases of the spacecraft’s rotation. With the transformation to an inertial frame, we factor out an integral over the spin phase, which we compute independently of the model distribution function. With the obtained spin-phase-averaged response function, the integration over the distribution function has the same computational complexity as the previously used method but is more accurate. In the future, our method might also be adapted to help analyze aspects of the Solar Wind and Pickup Ion (SWAPI) observations from the Interstellar Mapping and Acceleration Probe (IMAP) \citep{2018SSRv..214..116M}.

To determine the efficiency of the instrument for helium ions relative to that for protons, we compare the abundance of alpha particles in the solar wind at New Horizons with the 1 au data. We account for the propagation time from 1 au to New Horizons and for charge transfer processes, turning core solar wind protons into pickup ions, and partial neutralization of alpha particles. Finally, we average the 1 au and New Horizons observations over the solar rotation. The comparison shows that the relative efficiency is $73\% \pm 5\%$ higher for helium ions compared to that for protons, i.e., more than the previously assumed 50\% increase \citep[e.g.,][]{swaczyna_he_2019}. Unfortunately, the method inevitably relies on the calibration of the 1 au alpha-to-proton density ratio in the OMNI database, which is also uncertain. 

The main objective of the paper is to find the photoionization rates of ISN helium in the heliosphere independently from measurements on absolute calibration of the solar UV flux. After accounting for the increased efficiency for helium, we find that the SWAP-derived helium ionization rates are $43\% \pm 3\%$ higher than predicted by the photoionization model calibrated using measurements of the Sun UV spectrum and F10.7 flux as the proxy of its temporal evolution \citep{2020ApJ...897..179S}. This suggests that the level of solar UV radiation is higher than indicated by TIMED data. Accurate measurements of solar UV irradiance are essential, as it is the dominant energy source for heating the upper atmosphere \citep{2005JGRA..110.1312W} and plays a key role in global circulation models \citep{2008AdSpR..42..926Q, 2012JGRA..117.9307D}. Our result confirms the hypothesis of a higher ionization rate postulated by \citep{swaczyna_very_2022} based on the IBEX-Lo observations of ISN helium atoms. Observations from both IBEX-Lo and SWAP are consistent with $\sim$40\% more helium ionization. Moreover, we test an alternative hypothesis that the observed changes result from large-scale fluctuations of the ISN helium density. However, in this case, the SWAP and IBEX-Lo data predict opposite temporal gradients of the ISN density over corresponding periods, suggesting rejection of this hypothesis. Future observations from IBEX-Lo and IMAP-Lo will allow for verification of this conclusion. Furthermore, the IMAP-Lo’s pivot platform will allow for independent measurement of the ionization rate by direct sampling of direct and indirect ISN helium beam \citep{2023ApJS..265...24B}. 

Our conclusions suggest a more important role for helium in the heliosphere. With stronger ionization of helium ions, we expect more $\mathrm{He^+}$ PUIs and larger momentum transfer to the solar wind plasma. While many global heliosphere models neglect helium ions and atoms in the calculations, recent studies suggest a non-negligible impact of these populations \citep{2023ApJ...946...97F, 2024JPhCS2742a2011F}. Furthermore, the fluxes of helium energetic neutral atoms produced from the neutralization of $\mathrm{He^+}$ PUIs in the heliosheath should be stronger than previously predicted \citep{2013A&A...549A..76G,2017ApJ...840...75S}. Therefore, the results of this paper can also be tested by IMAP-Hi as ability to distinguish chemical elements in energetic neutral atom fluxes.

\begin{acknowledgments}
M.A. and P.S. are supported by a project financed by the Polish National Agency for Academic Exchange (NAWA) within the Polish Returns Programme (BPN/PPO/2022/1/00017) and the National Science Centre, Poland (2023/51/D/ST9/01261). M.B. acknowledges Polish National Science Centre grant 2023/51/B/ST9/01921. D.J.M. was supported by the IBEX (80NSSC18K0237) and IMAP (80GSFC19C0027) grants from NASA. For the purpose of open access, the authors have applied a CC-BY public copyright license to any Author Accepted Manuscript version arising from this submission.
\end{acknowledgments}

\begin{contribution}

M.A. made a formal analysis, wrote the software, and was responsible for writing and submitting the manuscript.
P.S. came up with the initial research concept, formulated research goals, designed the methodology, validated results, and edited the manuscript.
D.J.M. is the IBEX mission SWAP instrument principal investigator and provided insight on the SWAP efficiency for different species.
H.A.E. provided the solar wind data from SWAP fine sweeps. 
M.B. provided the photoionization rates data. 
All authors made constructive comments and suggestions, revised and accepted the manuscript.

\end{contribution}

\section*{Data availability}

SWAP calibrated data products from the New Horizons Mission and NH Kuiper Belt Extended Missions (KEM and KEM2) Data Sets were used for the analysis contained within this paper \citep{McComas_2017, McComas_2018, McComas_2019, McComas_2023, McComas_2024}
The SWAP data can be downloaded from: \url{https://pds-smallbodies.astro.umd.edu/}.
The OMNI 2 data can be downloaded from: \url{https://omniweb.gsfc.nasa.gov/ow.html}.

\software{The Python code used in this analysis is publicly available on Zenodo \url{https://doi.org/10.5281/zenodo.17379036}.}

\bibliography{bibliography.bib}{}
\bibliographystyle{aasjournal}

\end{document}